\documentstyle[prd,aps,floats,tighten,epsfig,graphicx,psfrag,here]{revtex}
\newcommand{\bda}{\begin{\displaymath}\begin{array}{rl}}
\newcommand{\eda}{\end{array}\end{displaymath}}
\newcommand{\be}{\begin{equation}}
\newcommand{\ee}{\end{equation}}
\newcommand{\bdm}{\begin{displaymath}}
\newcommand{\edm}{\end{displaymath}}
\newcommand{\bea}{\begin{eqnarray}}
\newcommand{\eea}{\end{eqnarray}}
\newcommand{\no}{\nonumber \\}
\newcommand{\fs}{\; \; .}
\newcommand{\co}{\; \; ,}

\newcommand{\ind}{\scriptscriptstyle}

\newcommand{\ubar}{\overline{\rule[0.42em]{0.4em}{0em}}\hspace{-0.5em}u}

\newcommand{\lbar}{\bar{\ell}}
\newcommand{\lsim}{\,\raisebox{-0.3em}{$\stackrel{\raisebox{-0.1em}{$<$}}{\sim}$
}\,}

\newcommand{\lvac}{\langle 0|\,}
\newcommand{\rvac}{\,|0\rangle}
\newcommand{\rs}{\langle r^2\rangle\rule[-0.2em]{0em}{0em}_s}
\begin{document}

\title{The quark condensate from \boldmath{$K_{e_4}$} decays }


\vspace{0.5cm}
\author{G.~Colangelo}\address{Institute for Theoretical Physics, University of 
Z\"urich, Winterthurerstr. 190, CH-8057 Z\"urich, Switzerland} 
\author{J.~Gasser and H.~Leutwyler}
\address{Institute for Theoretical Physics, University of 
Bern, Sidlerstr. 5, CH-3012 Bern, Switzerland}
\twocolumn[\hsize\textwidth\columnwidth\hsize\csname@twocolumnfalse\endcsname
\maketitle 

\begin{abstract}
We show that, independently of the size of the quark condensate, chiral
symmetry correlates the two
$S$--wave $\pi\pi$ scattering lengths. In view of this constraint, the
new precision data on $K_{e_4}$ decay allow a remarkably accurate
determination of these quantities. The result confirms the
hypothesis that the quark condensate is the leading order parameter.\\
\end{abstract}

]\narrowtext  

\thispagestyle{empty}

Since the masses of the two lightest quarks are very small, the
Hamiltonian of QCD is almost exactly invariant under the group 
SU(2)$_{\ind R}\times$SU(2)$_{\ind L}$ of chiral rotations. 
On phenomenological grounds, it is known that this
symmetry is spontaneously broken, the pions playing the role of the
corresponding Goldstone bosons \cite{Nambu}. 
If the symmetry was exact, the pions would be massless. According to
Gell-Mann, Oakes and Renner \cite{GMOR}, the square of the pion mass is
proportional to the product of the quark masses and the quark condensate:
\bea\label{eq:GMOR} 
M^2_\pi\simeq\frac{1}{F_\pi^2}\times(m_u+m_d)\times |\lvac\, \ubar u\rvac|
\fs\eea
The factor of proportionality is given by the pion decay constant
$F_\pi$. The term $m_u+m_d$ measures the explicit
breaking of chiral symmetry,  
while the quark condensate $\lvac\, \ubar u\rvac$
is a measure of the spontaneous symmetry breaking: It may be viewed as an
order parameter and plays a role analogous to the spontaneous 
magnetization of a magnet.  

The approximate
validity of the relation (\ref{eq:GMOR}) was put to question by Stern and
collaborators \cite{KMSF}, who pointed out that there is no experimental
evidence for the quark condensate to be different from zero. Indeed,
the dynamics of the ground state of QCD is not understood -- it could resemble
the one of an antiferromagnet, where, for dynamical reasons, the
most natural candidate for an order parameter, the magnetization, 
happens to vanish. 
What can be shown from first principles is only that (a) the expansion of
$M_\pi^2$ in powers of the quark masses starts with a linear
term,
\bea\label{eq:Mexp} M_\pi^2&=&M^2-\frac{\lbar_3}{32\pi^2 F^2}\,M^4
+O(M^6)\co\\
M^2&=&(m_u+m_d)\,B\co \nonumber\eea
and (b) the coefficient $B$ of the linear term is given by the value of 
$|\lvac\, \ubar u\rvac|/F_\pi^2$ in the limit $m_u,m_d\rightarrow 0$. 
The quantity $\lbar_3$ is one of the coupling constants occurring in the
effective Lagrangian at order $p^4$. The symmetry does not determine its
size. The crude estimates 
underlying the standard version of Chiral Perturbation Theory \cite{GL 1984}
indicate values in the range $0<\lbar_3<5$.
The term of order $M^4$ is then very small compared to the one of order $M^2$,
so that the Gell-Mann-Oakes-Renner formula is obeyed very well.
Stern and collaborators investigate the more general framework, referred to as
generalized CHPT, where
arbitrarily large values of $\lbar_3$ are considered. The
quartic term in eq.~(\ref{eq:Mexp}) can then take values comparable to the
``leading'', quadratic one. If so, the dependence of $M_\pi^2$ on the quark 
masses would fail to be approximately linear, even for values of $m_u$ and 
$m_d$ of the order of a few MeV. A different bookkeeping for the terms 
occurring in the chiral perturbation series is then needed \cite{KMSF} 
-- the standard chiral power counting is adequate only if $\lbar_3$ is 
not too large.

The purpose of the present note is to show that (a) in the generalized
scenario, the low energy structure is controlled by a single parameter, 
(b) this parameter can be determined  on the
basis of the $K_{e_4}$ data taken recently at Brookhaven
\cite{Brookhaven,Truol} and (c) the result beautifully confirms the
Gell-Mann-Oakes-Renner formula. 

The following analysis relies on the fact that the low energy properties of
the pions are controlled by two parameters: the $S$--wave
scattering lengths $a_0^0$, $a_0^2$. If these are given, the Roy 
equations \cite{Roy} allow us to calculate the scattering
amplitude in terms of the absorptive parts above 800 MeV and the available 
experimental information about the latter suffices to evaluate the relevant
dispersion integrals, to within small uncertainties \cite{ACGL}. 

Weinberg's low energy theorem \cite{Weinberg 1966} predicts the two 
scattering lengths in terms of the pion decay constant, so that the scattering
amplitude is then fully determined. The prediction is of limited accuracy,
because it only holds to leading order of an expansion in powers of the quark
masses $m_u$ and $m_d$. At first nonleading order of the expansion in powers
of momenta and quark masses,  the scattering amplitude
can be expressed in terms of $F_\pi$, $M_\pi$ and the 
coupling constants $\ell_1,\ldots\,,\ell_4$ that occur in the derivative
expansion of the effective Lagrangian at order $p^4$ (throughout, we ignore  
isospin breaking effects and work with $m_u=m_d=m$). The terms $\ell_1$ and
$\ell_2$ manifest themselves in the energy dependence 
of the scattering amplitude and can thus be determined phenomenologically. 
The term $\ell_3$ was mentioned above -- the range considered for this
coupling constant makes the difference between standard and generalized CHPT.  
Finally, $\ell_4$ is related to the slope of the
scalar form factor, which is known rather accurately from dispersion theory:
$\rs=0.61\pm 0.04\,\mbox{fm}^2$ \cite{CGL,Moussallam:2000aq,Descotes:2000ct}.  

As pointed out long ago \cite{GL 1983}, there is a low energy theorem
that relates the $S$--wave scattering lengths to the scalar radius:
\bea\label{eq:one loop} &&2a_0^0-5a_0^2=\\&&
\frac{3\,M_\pi^2}{4\pi F_\pi^2}\left\{1+
\frac{1}{3}\,M_\pi^2 \rs+\frac{41\,M_\pi^2}{192 \,\pi^2 F_\pi^2}\right\}
 + 
O(m^3)\fs\nonumber\eea
The theorem shows that the first order correction to the Weinberg formula 
for this particular combination of scattering lengths is determined by $\rs$. 
It correlates the two scattering lengths, irrespectively
of the numerical value of $\ell_3$: The correlation holds 
both in standard and generalized CHPT.

The corrections occurring in eq.~(\ref{eq:one loop})
at order $m^3$ are also known \cite{BCEGS}.
In the following, we analyze the correlation at that level of precision, using
the method described in ref.~\cite{CGL} -- except that we now treat the 
coupling constant $\ell_3$ as a free parameter. 
For the symmetry breaking
couplings entering the effective Lagrangian at order $p^6$, we assume
that the estimates given in ref.~\cite{BCT} are valid within a factor of 2
(see \cite{opus2} for a detailed error analysis).
The experimental input used for the Roy equations is taken from
ref.~\cite{ACGL}. Up to the noise attached to these ingredients,
the Roy equations then determine the scattering amplitude 
as a function of the parameter $\ell_3$.
In particular, we may calculate $a_0^0$ and $a_0^2$ as functions of 
$\ell_3$. The result is  well described by a parabola:
\bea\label{eq:corral3} a_0^0&=& 
.225-1.6\cdot 10^{-3}\,\lbar_3-1.3 \cdot 10^{-5}\,(\lbar_3)^{2}\co\\
a_0^2&=& -.0434-3.6\cdot 10^{-4}\,\lbar_3-4.3\cdot 10^{-6}\,
(\lbar_3)^{2}\fs\nonumber\eea

Eliminating the parameter $\ell_3$, we obtain the following correlation 
between $a_0^2$ and $a_0^0$: 
\bea\label{eq:corra0a2} 
a_0^2&=&-.0444\pm .0008 +.236\, (a_0^0-.22)\\&&-0.61\,
(a_0^0-.22)^2 -9.9\, (a_0^0-.22)^3\fs\nonumber\eea 
The error given accounts for the various sources of uncertainty in our 
input. The relation is indicated
\begin{figure}[thb]
\leavevmode
\begin{center}
\includegraphics[width=7.5cm]{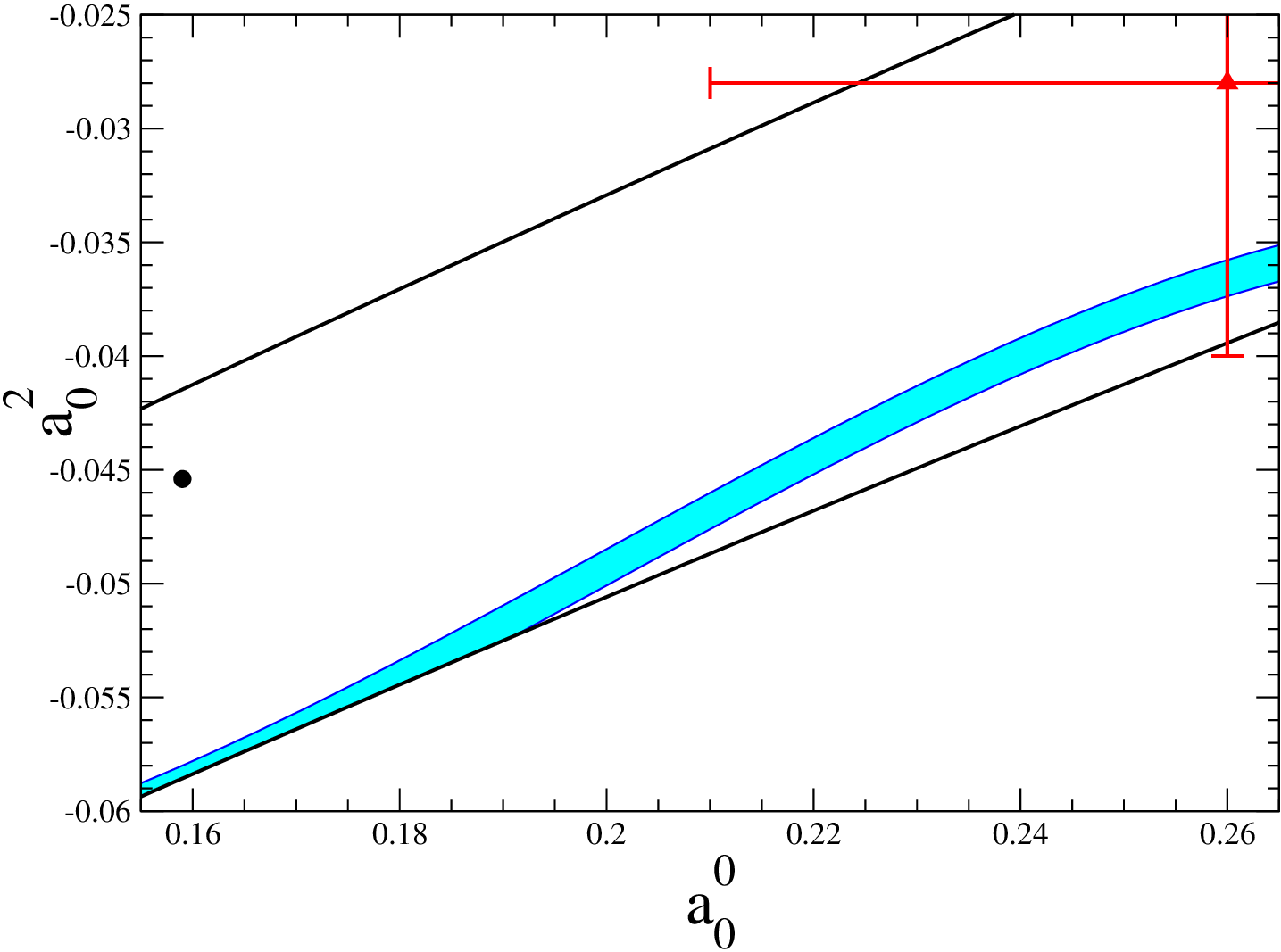}
\end{center}
\caption{\label{fig:aellipse} $S$--wave scattering
  lengths. The Roy
  equations only admit solutions in the ``universal band'',  
  spanned by the two tilted lines. The correlation in eq.~(\ref{eq:corra0a2}) 
  constrains the values to the narrow shaded strip.
  The full circle indicates Weinberg's leading order
  result, while the triangle with error bars shows the phenomenological
  range permitted by the old data, $a_0^0=0.26\pm 0.05$,
  $a_0^2=-0.028\pm0.012$ \protect\cite{Froggatt:1977hu}. }
\end{figure}
in fig.~\ref{fig:aellipse}: The values of $a_0^0$ and $a_0^2$ are
constrained to a narrow strip that runs along the lower edge of the 
universal band, which is indicated by the tilted straight lines.
As discussed in ref.~\cite{ACGL}, a qualitatively similar
correlation also  results from the Olsson sum rule \cite{olsson sum rule}
-- the two conditions are perfectly compatible, but the one above is
considerably more stringent. 

The analysis of the final state distribution observed in the decay 
$K\rightarrow \pi\pi e \nu$ yields a measurement of the phase difference
$\delta(s)\equiv\delta_0^0(s)-\delta_1^1(s)$, for
$4M_\pi^2<s<M_K^2$. At those energies,
$\delta(s)$ is dominated by the contribution
$\propto a_0^0$ from the $S$--wave scattering length. 
The correlation between $a_0^2$ and $a_0^0$ allows us to correct for the 
higher order terms of the threshold expansion and to express the phase 
difference in terms of $a_0^0$ and $q$, where $q$ is the c.m.~momentum
in units of $M_\pi$, $s= 4M_\pi^2(1+q^2)$.
In the region of
interest ($q<1$, $0.18<a_0^0<0.26$), the prediction reads 
\bea\label{eq:delta(s)} &&\delta_0^0-\delta_1^1=
\frac{q}{\sqrt{1+q^2}}\,( a_0^0+q^2\,b+q^4\,c+q^6\,d)\pm e\\
&&b= .2527+.151\,\Delta a_0^0+1.14\,(\Delta a_0^0)^2 +
35.5\,(\Delta a_0^0)^3\co\no
&&c=.0063-.145\,\Delta a_0^0\co\hspace{2em}
d=-.0096\co\nonumber\eea
with $\Delta a_0^0=a_0^0-.22$.
The uncertainty in this relation mainly stems from the experimental input used
in the Roy equations and is not sensitive to $a_0^0$:
\bea\label{eq:errordelta} e= .0035 \,q^3+.0015\,q^5\fs\eea
\begin{figure}[thb]
\psfrag{0.18}{\hspace{0.5em}\raisebox{-0.2em}{$0.18$}}
\psfrag{0.22}{\hspace{0.5em}\raisebox{-0.2em}{$0.22$}}
\psfrag{0.26}{\hspace{0.5em}\raisebox{-0.2em}{$0.26$}}
\leavevmode

\centering
\includegraphics[width=8cm]{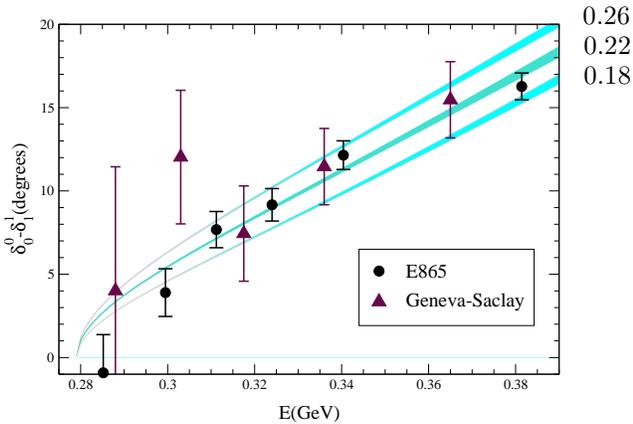}
\caption{\label{fig:deltaKl4} Phase relevant for the decay $K\rightarrow
  \pi\pi e\nu$. The three bands correspond to the 
three indicated values of the $S$--wave scattering length $a_0^0$. The
uncertainties are dominated by those from the experimental input used
in the Roy equations. The triangles are the data points of
ref.~\protect\cite{rosselet}, while the full circles represent the 
preliminary E865 results \protect\cite{Truol}.}
\end{figure}
The prediction (\ref{eq:delta(s)}) is illustrated
in fig.~\ref{fig:deltaKl4}, where the energy dependence of  
the phase difference is shown for $a_0^0=0.18$, $0.22$ and
$0.26$. The width of the corresponding bands indicates the uncertainties,
which according to (\ref{eq:errordelta})
grow in proportion to $q^3$ -- in the range shown, they
amount to less than a third of a degree. 

The figure shows that the data of ref.~\cite{rosselet} barely 
distinguish between the three values of $a_0^0$ shown.  
The preliminary results of the E865 experiment at Brookhaven \cite{Truol} are
significantly more precise, however. The best fit to these data is obtained
for $a_0^0=.218$, with $\chi^2= 5.7$ for 5 degrees of freedom. This
beautifully confirms the very sharp predictions obtained on the basis of  
standard CHPT: $a_0^0=.220\pm0.005$, $a_0^2=-.0444\pm.0010$ 
\cite{CGL,ABT}. There is a marginal
problem only with the bin of lowest energy: The corresponding scattering
lengths are outside the region
where the Roy equations admit solutions. In view of the
experimental uncertainties attached to that point, this discrepancy is without
significance: The difference between the central experimental value and 
the prediction amounts to $1\frac{1}{2}$ 
standard deviations. Note also that the old data are perfectly consistent with
the new ones: The overall fit
yields $a_0^2=.221$ with $\chi^2= 8.3$ for 10 degrees of freedom

\begin{figure}[t]
\leavevmode
\centering
\includegraphics[width=8cm]{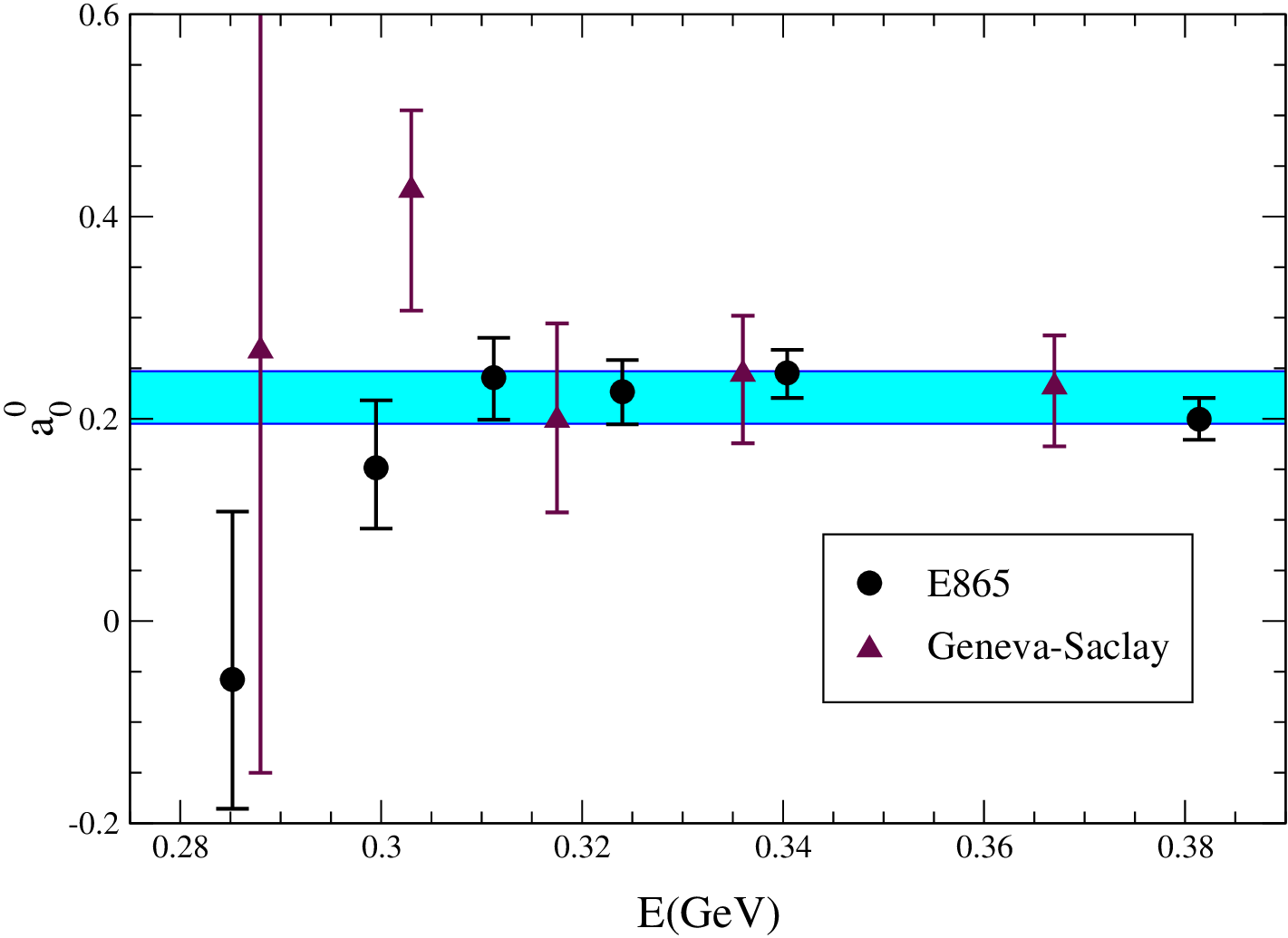}
\caption{\label{fig:aKe4}  $K_{e_4}$ data on the scattering length $a_0^0$.  
The triangles are the data points of
ref.~\protect\cite{rosselet}, while the full circles represent the preliminary
E865 results \protect\cite{Truol}. The horizontal band indicates the 
statistical average of the 11 values for $a_0^0$ shown in the figure.}
\end{figure}

The relation  (\ref{eq:delta(s)}) can be inverted, so
that each one of the values found for the phase difference yields
a measurement of the scattering length $a_0^0$.  
The result is shown in fig.~\ref{fig:aKe4}.
The experimental errors are remarkably small. It is not unproblematic,
however, to treat the data collected in the 
different bins as statistically independent: In the presence of correlations,
this procedure underestimates the actual uncertainties. Also, since the phase
difference rapidly rises with the energy, the binning procedure may introduce
further uncertainties. To be on the conservative side, we estimate the
uncertainties by using the 95\% confidence limit, where we obtain
$a_0^0=.221\pm .026$. 
For the final data analysis, we refer to a forthcoming paper by the E685
collaboration.

We may translate the result into an estimate for the magnitude 
of the coupling constant $\lbar_3$. In this language, the above conclusion for
the value of $a_0^0$ corresponds to $|\lbar_3|\,\lsim\, 16$.
Although this is a coarse estimate, it implies that the
Gell-Mann-Oakes-Renner relation does represent a decent approximation: 
More than 94\% of the pion mass must stem from the quark 
condensate.

\acknowledgments We are indebted to S.~Pislak and P.~Tru\"ol for providing
us with preliminary data of the E865 collaboration. 
This work was supported in part by the Swiss National Science
Foundation, and by TMR, BBW-Contract No. 97.0131  and  EC-Contract
No. ERBFMRX-CT980169 (EURODA$\Phi$NE).


\begin{thebibliography}{99}
\bibitem{Nambu}
Y.~Nambu,
Phys.\ Rev.\ Lett.\ {\bf 4} (1960) 380;
Phys.\ Rev.\ {\bf 117} (1960) 648.

\bibitem{GMOR}
M.~Gell-Mann, R.~J.~Oakes and B.~Renner,
Phys.\ Rev.\ {\bf 175} (1968) 2195.

\bibitem{KMSF}

M.~Knecht, B.~Moussallam, J.~Stern and N.~H.~Fuchs,
Nucl.\ Phys.\  {\bf B457} (1995) 513
[hep-ph/9507319];
ibid. {\bf B471} (1996) 445
[hep-ph/9512404].

\bibitem{GL 1984}
J.~Gasser and H.~Leutwyler,
Annals Phys.\  {\bf 158} (1984) 142.

\bibitem{Brookhaven}
S.~Pislak {\it et al.}, ``A new measurement of $K^+\rightarrow
\pi^+\pi^-e^+\nu$ ($K_{e4}$)'', talk at  Laboratori Nazionali di 
  Frascati, June 22, 2000.

\bibitem{Truol}P.~Truol {\it et al.}  [E865 Collaboration],
hep-ex/0012012.

\bibitem{Roy}
S.~M.~Roy,
Phys.\ Lett.\  {\bf B36} (1971) 353.

\bibitem{ACGL}
B.~Ananthanarayan, G.~Colangelo, J.~Gasser and H.~Leutwyler,
Phys. Rept. (in press) 
[hep-ph/0005297].

\bibitem{Weinberg 1966}
S.~Weinberg,
Phys.\ Rev.\ Lett.\  {\bf 17} (1966) 616.


\bibitem{CGL}
G.~Colangelo, J.~Gasser and H.~Leutwyler,
Phys.\ Lett.\  {\bf B488} (2000) 261
[hep-ph/0007112].

\bibitem{Moussallam:2000aq}
B.~Moussallam,
Eur.\ Phys.\ J.\ C {\bf 14} (2000) 111
[hep-ph/9909292].

\bibitem{Descotes:2000ct}
S.~Descotes,
hep-ph/0012221.

\bibitem{GL 1983}
J.~Gasser and H.~Leutwyler,
Phys.\ Lett.\  {\bf B125} (1983) 325.

\bibitem{BCEGS}
J.~Bijnens, G.~Colangelo, G.~Ecker, J.~Gasser and M.~E.~Sainio,
Phys.\ Lett.\  {\bf B374} (1996) 210
[hep-ph/9511397];
Nucl.\ Phys.\  {\bf B508} (1997) 263
[hep-ph/9707291];
ibid. {\bf B517} (1998) 639 (E). 

\bibitem{BCT}
J.~Bijnens, G.~Colangelo and P.~Talavera,
JHEP {\bf 9805} (1998) 014
[hep-ph/9805389].

\bibitem{opus2} G.~Colangelo, J.~Gasser and H.~Leutwyler,
{\em $\pi\pi$ scattering}, in preparation.

\bibitem{Froggatt:1977hu}
C.~D.~Froggatt and J.~L.~Petersen,
Nucl.\ Phys.\ B {\bf 129} (1977) 89;\\
M.~M.~Nagels {\it et al.},
Nucl.\ Phys.\ B {\bf 147} (1979) 189.

\bibitem{olsson sum rule}
M.~G.~Olsson, 
Phys.\ Rev.\ {\bf 162} (1967) 1338.

\bibitem{rosselet}
L.~Rosselet {\it et al.},
Phys.\ Rev.\  {\bf D15} (1977) 574.

\bibitem{ABT}
G.~Amoros, J.~Bijnens and P.~Talavera
[hep-ph/0003258].


\end{thebibliography}
\end{document}